\documentclass[a4paper,11pt]{article}
\pdfoutput=1 
\usepackage{jheppub} 
\usepackage[T1]{fontenc} 
\usepackage{multirow}
\usepackage[numbers,sort&compress]{natbib}
\usepackage{amsmath}
\usepackage{amssymb}
\usepackage[normalem]{ulem}
\usepackage{graphicx}
\usepackage{dcolumn}
\usepackage{bm}
\usepackage{color}
\usepackage{float}
\usepackage{multirow}
\usepackage{multicol}
\usepackage{subfigure}

\newcommand{\f}{\frac}
\newcommand{\lt}{\left}
\newcommand{\n}{\nonumber}
\newcommand{\p}{\partial}
\newcommand{\dd}{{\rm d}}
\newcommand{\rt}{\right}

\newcommand{\arxph}[1]{\href{http://arxiv.org/abs/#1}{{\ttfamily arXiv:#1[physics]}}}
\newcommand{\arxgr}[1]{\href{http://arxiv.org/abs/#1}{{\ttfamily arXiv:#1[gr-qc]}}}
\newcommand{\arxth}[1]{\href{http://arxiv.org/abs/#1}{{\ttfamily arXiv:#1[hep-th]}}}

\newcommand{\Arxgr}[1]{\href{http://arxiv.org/abs/gr-qc/#1}{{\ttfamily arXiv:#1[gr-qc]}}}
\newcommand{\Arxth}[1]{\href{http://arxiv.org/abs/hep-th/#1}{{\ttfamily arXiv:#1[hep-th]}}}

\title{\boldmath The Joule--Thomson and Joule--Thomson-like effects of the black holes in a cavity}
\author[a,1]{Nan Li, \note{Corresponding author.}}
\author[a]{Jin-Yu Li,}
\author[a]{and Bing-Yu Su}
\affiliation[a]{Department of Physics, College of Sciences, Northeastern University, Shenyang 110819, China
}
\emailAdd{linan@mail.neu.edu.cn}

\abstract{When a black hole is enclosed in a cavity in asymptotically flat space, an effective volume can be introduced, and an effective pressure can be further defined as its conjugate variable. By this means, an extended phase space is constructed in a cavity, which resembles that in the anti-de Sitter (AdS) space in many aspects. However, there are still some notable dissimilarities simultaneously. In this work, the Joule--Thomson (JT) effect of the black holes, widely discussed in the AdS space as an isenthalpic (constant-mass) process, is shown to only have cooling region in a cavity. On the contrary, in a constant-thermal-energy process (the JT-like effect), there is only heating region in a cavity. Altogether, different from the AdS case, there is no inversion temperature or inversion curve in a cavity. Our work reveals the subtle discrepancy between the two different extended phase spaces that is sensitive to the specific boundary conditions.}

\begin{document}
\maketitle
\flushbottom

\section{Introduction} \label{sec:intro}

In recent years, black hole thermodynamics in the so-called ``extended phase space" has aroused increasing research interest. The fundamental motivations are as follows. Albeit black hole thermodynamics exhibits remarkable similarity to traditional thermodynamics \cite{law}, there remain two distinct differences in between. First, the usual $p$--$V$ term is absent in the first law of black hole thermodynamics. Second, a black hole in asymptotically flat space has negative heat capacity and is thus thermodynamically unstable, so it cannot maintain thermal equilibrium with the environment and will eventually evaporate via the Hawking radiation \cite{rad}. Altogether, a black hole is still somehow different from a traditional thermodynamic system (e.g., a $p$--$V$--$T$ system).

Actually, both of the above two discrepancies can be dealt with to be consistent with traditional thermodynamics. The basic idea consists of two steps. First, one needs to impose appropriate boundary conditions instead of the asymptotically flat one. Second, an effective pressure and an effective volume should be introduced to the black hole system in order to restore the $p$--$V$ term. This means that two new dimensions are added into the phase space of the black hole, so that the analogy between black hole thermodynamics and traditional thermodynamics becomes more complete.

In general, there are two natural ways to construct the extended phase space. The first is to confine the black hole in the anti-de Sitter (AdS) space, and an effective pressure $p$ can be introduced as \cite{Kastor, Dolan}
\begin{align}
p=-\f{\Lambda}{8\pi}, \n
\end{align}
where $\Lambda$ is the negative cosmological constant. If $\Lambda$ is running, $p$ turns to be a thermodynamic variable rather than a fixed background, so an effective volume $V$ of the black hole can then be defined as its conjugate variable as $V=({\p M}/{\p p})_S$, with $M$ being the black hole mass (i.e., the enthalpy) \cite{KM}. Furthermore, in the AdS space, there are two branches of black hole solutions: a stable large black hole with larger event horizon radius and positive heat capacity and an unstable small black hole on the contrary. The large black hole has an equation of state (EoS) similar to that of a non-ideal fluid and thus exhibits rich thermodynamic behaviors, such as various phase transitions and abundant critical phenomena \cite{rev}.

The second choice of the extended phase space is to enclose the black hole in a cavity in asymptotically flat space, on the wall of which the metric is fixed (i.e., with the Dirichlet boundary condition) \cite{York, Braden, Parentani:1994wr, Gregory:2001bd, Carlip, Lundgren, Akbar:2004ke, Emparan:2012be, Dolansss, 27, Ponglertsakul1, 28, 29, Ponglertsakul2, Sanchis-Gualhhh, 31, 30, Dias11, Dias12, 35, Wang:2019kxp, 36, Wang:2019urm, Wang:2019cax, 37, Wang:2020osg, Tzikas, zwb, Yao:2021zid, ElMoumni:2021woq, Su:2021jto, Wang:2021llu, Feng:2021zsq, Huang:2021eby, Cao:2022hmd, Huang:2022agr}. In this way, the wall acts as a reflecting boundary against the Hawking radiation and thus stabilizes the black hole. Hence, the function of the cavity is similar to the AdS space, as the gravitational potentials increase at large distances in both of these two cases, only with different boundary conditions. Inspired by this observation, the authors of Ref. \cite{Wang} constructed a new extended phase space in a cavity, with an effective volume $V$ of the black hole introduced as
\begin{align}
V=\f{4\pi}{3}r^3_B, \n
\end{align}
where $r_B$ is the cavity radius. If $r_B$ is allowed to vary, $V$ becomes a thermodynamic variable. Furthermore, if we demand the validity of the first law of black hole thermodynamics, an effective pressure $p$ can be accordingly defined as the conjugate variable of $V$ as $p=-({\p E}/{\p V})_{S}$, with $E$ being the thermal energy of the black hole. Compared with black hole thermodynamics in the AdS space, which has received extensive studies in the past decade, the relevant research in the cavity case is just starting.

It is very interesting to point out that the order of the introduction of the $p$--$V$ term is opposite in these two different extended phase spaces. Consequently, besides the expected similarities (e.g., the same critical exponents \cite{KM, Wang}), there should also be important differences at the same time (e.g., the global phase structure \cite{zwb} and the dual relation \cite{Su:2021jto} in the Hawking--Page phase transition). Moreover, the $p$--$V$ terms in the first law of black hole thermodynamics are $V\,\dd p$ in the AdS case but $-p\,\dd V$ in the cavity case, so the first laws contain different thermodynamic variables: the enthalpy $M$ for the former but the thermal energy $E$ for the latter. This discrepancy will lead to an essential difference in another typical thermodynamic process, the Joule--Thomson (JT) effect, between the AdS and cavity cases, and the relevant exploration is the topic of our present work.

Since a large black hole in the extended phase space possesses an EoS like that of a non-ideal fluid, it may experience a temperature change in the JT effect. In the AdS space, this effect was first studied for the charged black hole \cite{okcu1}, then for the rotating one \cite{okcu2}, and finally for the most general charged and rotating one \cite{Zhao:2018kpz}. These works indicated that, different from the van der Waals-like fluid, the black hole in the AdS space has only a minimum inversion temperature, without a maximum one. The successive studies in various extended theories of gravity further confirmed these conclusions \cite{Ghaffarnejad:2018exz, Mo:2018rgq, Chabab:2018zix, Mo:2018qkt, Lan:2018nnp, Kuang:2018goo, Cisterna:2018jqg, Haldar:2018cks, Li:2019jcd, MahdavianYekta:2019dwf, Nam:2019zyk, Lan:2019kak, Sadeghi:2020bon, Rajani:2020mdw, Nam:2020gud, Guo:2020zcr, Hegde:2020xlv, Guo:2020qxy, Bi:2020vcg, Silva:2021qtw, Mirza:2021kvi, Zhang:2021raw, Liang:2021elg, Liang:2021xny, Graca:2021izb, Yin:2021akt, Zhang:2021kha, Biswas:2021uop, Graca:2021ker, Meng:2021cgb, Abdusattar:2021wfv, Xing:2021gpn, Zarepour:2021xmz, Assrary:2022uiu, Kruglov:2022lnc, Yin:2022mcy, Barrientos:2022uit, Hui:2022ubj}, and the basic reason is the slight difference between the EoS of the black hole and the van der Waals-like fluid.

However, the JT effect of the black holes in a cavity has not been thoroughly explored in the current literature. In Ref. \cite{Cao:2022hmd}, this issue was briefly touched, and it was claimed that there is only cooling region. Nevertheless, there still remain two areas needed for necessary improvements. One is that the authors merely illustrated their conclusion graphically, without detailed mathematical proof. The other but more fundamental one is that a naive isenthalpic process has no intrinsic physical essence for the black holes in a cavity. This is because it is the thermal energy $E$ that is formulated in the first law of black hole thermodynamics in a cavity, but not the enthalpy $M$ in the AdS space. Consequently, it is no longer suitable to consider the ordinary JT effect for the black holes in a cavity, but a JT-like effect (i.e., a constant-thermal-energy process rather than an isenthalpic one) is more meaningful for the relevant discussions. This is the aim of our present work, and we will rigorously prove that, even in this new process, there is only heating region in the JT-like effect. In all, there is no inversion temperature or inversion curve in both the JT and JT-like effects of the black holes in a cavity. These properties evidently differ from those in the AdS space and indicate the notable discrepancy between different extended phase spaces that is sensitive to the specific boundary conditions.

This paper is organized as follows. In Sec. \ref{sec:cavity}, we list the thermodynamic properties of the black holes in a cavity. The JT effect of the black holes as an isenthalpic process is studied in Sec. \ref{sec:JT}. Next, in Sec. \ref{sec:JTlike}, we generalize our investigations to the JT-like effect as a constant-thermal-energy process. We conclude in Sec. \ref{sec:con}. We work in the natural system of units by setting $c=G_{\rm N}=\hbar=k_{\rm B}=1$.

\section{Thermodynamics of the black holes in a cavity} \label{sec:cavity}

In this section, we first outline black hole thermodynamics in the extended phase space in a cavity and then explain the JT the JT-like effects in detail.

\subsection{Extended phase space in a cavity} \label{sec:excavity}

In four-dimensional space-time, the action of the Einstein--Maxwell gravity is
\begin{align}
I=\f{1}{16\pi}\int\dd^4x\,\sqrt{-g}(R-F_{\mu\nu}F^{\mu\nu}), \n
\end{align}
where $R$ is the Ricci scalar, and $F_{\mu\nu}$ is the electromagnetic field tensor. The metric of the charged black hole reads
\begin{align}
\dd s^2=-f(r)\,\dd t^2+\f{\dd r^2}{f(r)}+r^2\,\dd\theta^2+r^2\sin^2\theta\,\dd\phi^2,\n
\end{align}
where $f(r)=1-2M/r+Q^2/r^2$, and $M$ and $Q$ are the mass and electric charge of the black hole, with the electric potential being $A=A_t(r)\,\dd t=-Q\,\dd t/r$. The event horizon radius $r_+$ is determined as the largest root of $f(r_+)=0$, and we have $r_+=M+\sqrt{M^2-Q^2}$. To avoid naked singularity, $r_+$ must be positive, and this constraint requires $M>Q$, so $r_+>Q$, with $Q$ being the event horizon radius of the extremal black hole. In terms of $r_+$, $f(r)$ can be reexpressed as
\begin{align}
f(r)=\lt(1-\f{r_+}{r}\rt)\lt(1-\f{Q^2}{r_+r}\rt). \n
\end{align}
Furthermore, the Bekenstein--Hawking entropy is one quarter of the event horizon area,
\begin{align}
S=\pi r_+^2, \n
\end{align}
and the Hawking temperature can be calculated as
\begin{align}
T_{\rm H}=\f{f'(r_+)}{4\pi}=\f{1}{4\pi r_+}\lt(1-\f{Q^2}{r_+^2}\rt). \n
\end{align}

Next, we consider black hole thermodynamics in a cavity in asymptotically flat space. The cavity plays a role of a reservoir, with fixed temperature $T$ and electric potential $\Phi$ on its wall located at a radius $r_B$ \cite{York, Braden}. Here, we should emphasize an important point that the cavity radius $r_B$ should always be larger than the event horizon radius $r_+$. Therefore, we introduce
\begin{align}
x=\f{r_+}{r_B}, \quad y=\f{Q}{r_B}. \n
\end{align}
Considering $Q<r_+$, we definitely obtain
\begin{align}
0<y<x<1. \label{bds}
\end{align}
This inequality will have significant influences on the JT and JT-like effects of the charged black holes in a cavity, to be discussed in Secs. \ref{sec:JT} and \ref{sec:JTlike}.

In a cavity, the black hole temperature $T$ measured at $r_B$ is blue-shifted from the Hawking temperature $T_{\rm H}$ measured at infinity by a factor $1/\sqrt{f(r_B)}$ \cite{York, Braden},
\begin{align}
T =\f{T_{\rm H}}{\sqrt{f(r_B)}}=\f{1-\f{Q^2}{r_+^2}}{4\pi r_+\sqrt{\lt(1-\f{r_+}{r_B}\rt)\lt(1-\f{Q^2}{r_+r_B}\rt)}}. \label{T}
\end{align}
Moreover, the electric potential $\Phi$ measured at $r_B$ is
\begin{align}
\Phi=\f{A_t(r_B)-A_t(r_+)}{\sqrt{f(r_B)}}=\f{Q\lt(1-\f{r_+}{r_B}\rt)}{r_+\sqrt{\lt(1-\f{r_+}{r_B}\rt)\lt(1-\f{Q^2}{r_+r_B}\rt)}}. \n
\end{align}
Furthermore, the thermal energy $E$ of the charged black hole in a cavity can be achieved via the Euclidean action method \cite{Braden},
\begin{align}
E &= r_B\lt[1-\sqrt{f (r_B)}\rt]= r_B\lt[1-\sqrt{\lt(1-\f{r_+}{r_B}\rt)\lt(1-\f{Q^2}{r_+r_B}\rt)}\rt]. \label{E}
\end{align}

In the extended phase space in a cavity, the cavity radius $r_B$ is considered as a new thermodynamic variable, rather than a constant or a fixed background. In this way, an effective volume and an effective pressure appear in black hole thermodynamics. First, the effective volume $V$ of the black hole is introduced as \cite{Wang}
\begin{align}
V=\f{4\pi}{3}r^3_B. \label{V}
\end{align}
Then, we demand the validity of the first law of black hole thermodynamics in a cavity,
\begin{align}
\dd E = T\,\dd S - p\,\dd V+\Phi\,\dd Q. \label{firstc}
\end{align}
As a result, the effective pressure $p$ can be defined as the conjugate variable of $V$. From Eqs. (\ref{E}) and (\ref{V}), we have
\begin{align}
p&=-\lt(\f{\p E}{\p V}\rt)_{S,Q}=-\lt(\f{\p E}{\p r_B}\rt)_{r_+,Q}\f{\dd r_B}{\dd V}=\f{1}{4\pi r_B^2} \lt[\f{2-\f{r_+}{r_B}-\f{Q^2}{r_+r_B}}{2\sqrt{\lt(1-\f{r_+}{r_B}\rt)\lt(1-\f{Q^2}{r_+r_B}\rt)}}-1\rt]. \label{p}
\end{align}
By this means, the absent $p$--$V$ term can be restored in black hole thermodynamics. Moreover, a scaling argument easily leads to the Smarr relation \cite{Smarr} as $E=2TS-3pV+\Phi Q$ \cite{Wang}.

Altogether, the analogy between traditional thermodynamics and black hole thermodynamics in the extended phase space in a cavity is as complete as that in the AdS space. However, there exists an essential difference between them. In the AdS case, the first law of black hole thermodynamics is \cite{KM}
\begin{align}
\dd M = T\,\dd S + V\,\dd p+\Phi\,\dd Q. \label{firsta}
\end{align}
Therefore, the black hole mass $M$ should be interpreted as enthalpy in the AdS space, rather than thermal energy $E$, because the $p$--$V$ term is $V\,\dd p$ but not $-p\,\dd V$. This key point will cause significant discrepancies in the thermodynamic behaviors of the black holes in these two extended phase spaces.

Last, we stress that Eq. (\ref{p}) can also be viewed as the EoS of the black hole in a cavity. From Eqs. (\ref{T}) and (\ref{V}), solving $r_+$ and $r_B$ in terms of $T$ and $V$ and substituting them into Eq. (\ref{p}), we may obtain $p=p(V,T)$ in principle. Unfortunately, Eq. (\ref{T}) is an algebraic equation about $r_+$ of higher degree and is analytically unsolvable, so it is impossible to show the explicit form of $p(V,T)$. Therefore, we focus on two limiting cases: the large Schwarzschild black hole with $Q=0$ and the extremal black hole with $Q\to M$. In the limit $r_B\to\infty$, for the former, its EoS is
\begin{align}
p=\f{T}{2(3V/4\pi)^{1/3}}, \label{ddddd}
\end{align}
and for the latter, its EoS is
\begin{align}
p=\f{\pi Q^4T^2}{2(3V/4\pi)^{4/3}}. \label{eeeee}
\end{align}
These results indicate that the black holes in a cavity behave quite like the non-ideal fluids in traditional thermodynamics, so they may naturally experience the JT and JT-like effects to be discussed below. (For a general discussion on the EoS of the black hole in a cavity, see Ref. \cite{zwb}.)

\subsection{JT and JT-like effects}

The JT effect is an important issue in traditional thermodynamics, with widespread applications in thermal engineering. In this effect, a non-ideal fluid is forced through a valve or porous plug by a pressure difference, experiencing an adiabatic expansion with a temperature change. Albeit the JT effect is inherently irreversible, the enthalpies of the initial and final states of the fluid remain unchanged. Therefore, an adiabatic and isenthalpic process can be applied to calculate the temperature change.

Of course, there is no valve or porous plug in the Universe. However, in the extended phase spaces, a black hole has an EoS similar to that of a non-ideal fluid and may thus experience the JT effect. In this process, the temperature change is not caused by any usual heat exchange between the black hole and the environment, but is purely due to the work done by the variation of the effective pressure $p$ in the AdS space or by the variation of the effective volume $V$ in a cavity.

First, we review the JT effect in the AdS space, in which an isenthalpic process is basically a constant-mass process, as the black hole mass $M$ is in fact its enthalpy in this extended phase space. Therefore, the temperature change is encoded in the JT coefficient $\mu$,
\begin{align}
\mu=\lt(\f{\p T}{\p p}\rt)_{M,Q}=\f{1}{C_p}\lt[T\lt(\f{\p V}{\p T}\rt)_{p,Q}-V\rt], \label{JT}
\end{align}
where $C_p$ is the heat capacity at constant pressure. The sign of $\mu$ separates the $T$--$p$ plane into two regions: the cooling region with $\mu>0$ and the heating region with $\mu<0$, as pressure always decreases in the JT effect. Between these two regions where $\mu=0$, from Eq. (\ref{JT}), there is an inversion temperature $T_{\rm i}$ as the solution of
\begin{align}
\f TV=\lt(\f{\p T}{\p V}\rt)_{p,Q}, \label{Ti}
\end{align}
and the relevant relation $T_{\rm i}(p)$ defines the inversion curve.

Now, we turn to the cavity case. Naively, one may follow the same procedure as above, investigate an isenthalpic process, and calculate the JT coefficient. However, as we have emphasized in advance, there is an important difference between the AdS and cavity cases. From Eqs. (\ref{firstc}) and (\ref{firsta}), we clearly see that the thermodynamic variables are not the same. In the cavity case, it is the thermal energy $E$ that is in the first law, but not the enthalpy $M$ in the AdS case. Therefore, the JT effect as an isenthalpic process has no intrinsic physical essence for the black holes in a cavity. As a result, in this extended phase space, we should as well explore the corresponding constant-thermal-energy process for the black holes, and we dub this process a JT-like effect.

In the JT-like effect, we introduce a JT-like coefficient $\mu'$ as
\begin{align}
\mu'=\lt(\f{\p T}{\p V}\rt)_{E,Q}. \n
\end{align}
Using the relation between partial derivatives,
\begin{align}
\lt(\f{\p T}{\p V}\rt)_{E,Q}\lt(\f{\p V}{\p E}\rt)_{T,Q}\lt(\f{\p E}{\p T}\rt)_{V,Q}=-1, \n
\end{align}
we have
\begin{align}
\mu'=-\lt(\f{\p T}{\p E}\rt)_{V,Q}\lt(\f{\p E}{\p V}\rt)_{T,Q}=-\f{1}{C_V}\lt(\f{\p E}{\p V}\rt)_{T,Q}, \n
\end{align}
where $C_V$ is the heat capacity at constant volume. In Eq. (\ref{firstc}), changing the independent variables of $E$ from $(S,V,Q)$ to $(T,V,Q)$, we have
\begin{align}
\lt(\f{\p E}{\p V}\rt)_{T,Q}=T\lt(\f{\p S}{\p V}\rt)_{T,Q}-p=T\lt(\f{\p p}{\p T}\rt)_{T,Q}-p, \n
\end{align}
where the Maxwell relation is used. From the above two equations, we obtain the JT-like coefficient as
\begin{align}
\mu'=\f{1}{C_V}\lt[p-T\lt(\f{\p p}{\p T}\rt)_{V,Q}\rt]. \label{JTc}
\end{align}
This result clearly resembles the JT coefficient $\mu$ in Eq. (\ref{JT}), with the change of the variables from $p$ to $V$. Consequently, the sign of $\mu'$ also separates the $T$--$V$ plane into two regions: the cooling region with $\mu'>0$ and the heating region with $\mu'<0$. In between, the inversion curve is determined by setting $\mu'=0$ in Eq. (\ref{JTc}), and the inversion temperature $T_{\rm i}$ can be solved as the solution of
\begin{align}
\f Tp=\lt(\f{\p T}{\p p}\rt)_{V,Q}. \label{fz}
\end{align}
It is obvious to find that Eq. (\ref{fz}) is in parallel with Eq. (\ref{Ti}), showing the complementarity between these two extended phase spaces.

The explorations of the JT and JT-like effects of the black holes in a cavity constitute the main body of this paper. We will prove that, in both cases, there is no inversion temperature or inversion curve at all. The JT and JT-like coefficients will also be calculated in detail. In each section below, we start from the Schwarzschild black hole for clear physical comprehension and mathematical simplicity, and then generalize our results to the charged black hole and analyze the influence from the charge. The rotation of the black hole will be omitted, as it usually cannot contribute nontrivial physical insight, except increasing unnecessary mathematical tediousness.

\section{JT effect of the black holes in a cavity} \label{sec:JT}

In this section, we study the JT effect as an isenthalpic process for the black holes in a cavity. It will be shown that there is no inversion temperature or inversion curve, and the JT coefficient $\mu$ is positive definite.

\subsection{Inversion temperature}

From Eqs. (\ref{T}) and (\ref{V}), it is direct to obtain $T/V$ in Eq. (\ref{Ti}). However, the derivation of $(\p T/\p V)_{p,Q}$ is not straightforward anymore, with complicated derivatives of implicit functions evolved. From Eqs. (\ref{T}), (\ref{V}), and (\ref{p}), we have
\begin{align}
\lt(\f{\p T}{\p V}\rt)_{p,Q}=\lt(\f{\p T}{\p r_B}\rt)_{p,Q}\f{\dd r_B}{\dd V}
=\f{1}{4\pi r_B^2}\lt[\lt(\f{\p T}{\p r_B}\rt)_{r_+,Q} -\lt(\f{\p T}{\p r_+}\rt)_{r_B,Q}\f{(\p p/\p r_B)_{r_+,Q}}{(\p p/\p r_+)_{r_B,Q}}\rt]. \label{pTV}
\end{align}

First, for the Schwarzschild black hole, setting $Q=0$ in Eqs. (\ref{T}), (\ref{V}), (\ref{p}), and (\ref{pTV}), we have
\begin{align}
\f TV&=\f{3}{16\pi^2 r_B^4 x\sqrt{1-x}}, \label{jTV} \\
\lt(\f{\p T}{\p V}\rt)_{p,Q}&=-\f{7x^2-16x+8+4(3x-2)\sqrt{1-x}}{16\pi^2 r_B^4 x^3\sqrt{1-x}}, \label{jpTV}
\end{align}
with $x=r_+/r_B$. Substituting Eqs. (\ref{jTV}) and (\ref{jpTV}) into Eq. (\ref{Ti}), we easily find that the unique solution is $x=0$, so there is no inversion temperature or inversion curve for the Schwarzschild black hole in a cavity.

Second, for the charged black hole, the whole procedure remains the same, only with some mathematical complexity due to the charge. Again, we have
\begin{align}
\f TV&=\f{3(r_B^2x^2-Q^2)}{16\pi^2 r_B^6x^3\sqrt{f(r_B)}}, \label{jqTV} \\
\lt(\f{\p T}{\p V}\rt)_{p,Q}&=-\f{1}{16\pi^2 r_B^6 x^3 [r_B^4x^4-2r_B^2x^3Q^2+(2x-1)Q^4]\sqrt{f(r_B)}} \n\\
&\quad \times\lt\{r_B^6x^4\lt[7x^2-16x+8+4(3x-2)\sqrt{f(r_B)}\rt]\rt. \n\\
&\quad -r_B^4x^2\lt[2x^3+3x^2-32x+24+8(x^2+3x-3)\sqrt{f(r_B)}\rt]Q^2 \n\\
&\quad \lt. +r_B^2x\lt[4x^2-39x+32+4(6x-5)\sqrt{f(r_B)}\rt]Q^4+(14x-13)Q^6 \rt\}. \label{jqpTV}
\end{align}
If there were inversion temperature, equalizing Eqs. (\ref{jqTV}) and (\ref{jqpTV}) and remembering $y=Q/r_B$, we would need
\begin{align}
F(x,y)&=5 x^6 - 8 x^5 -4 x^5 y^2 + 4 x^4 - 3 x^4 y^2 + 16 x^3 y^2 + 8 x^3 y^4 - 12 x^2 y^2 - 21 x^2 y^4 \n\\
&\quad + 16 x y^4+ 4 x y^6 - 5 y^6 +2x (3 x^4 - 2 x^3 - 2 x^3 y^2 - 6 x^2 y^2 + 6 x y^2 + 6 x y^4- 5 y^4) \n\\
&\quad \times \sqrt{(1 - x) (1 - y^2/x)} = 0. \n
\end{align}

Unfortunately, this binary function $F(x,y)$ is positive definite under the constraint $0<y<x<1$ in Eq. (\ref{bds}), and can reach 0 only in the limit $y\to x$. Hence, there is no inversion temperature or inversion curve for the charged black hole in a cavity, either.

\subsection{JT coefficient}

In order to confirm the result in the previous subsection, we now further calculate the JT coefficient $\mu$. For this purpose, we start from the heat capacity at constant pressure $C_p$. After some algebra, we have
\begin{align}
\f{1}{C_p}&=\f 1T\lt(\f{\p T}{\p S}\rt)_{p,Q}=\f{(\p T/\p r_+)_{p,Q}}{T\,\dd S/\dd r_+} \n\\
&=\f{1}{2\pi r_+T}\lt[\lt(\f{\p T}{\p r_+}\rt)_{r_B,Q} -\lt(\f{\p T}{\p r_B}\rt)_{r_+,Q}\f{(\p p/\p r_+)_{r_B,Q}}{(\p p/\p r_B)_{r_+,Q}}\rt]. \label{Cp}
\end{align}

First, for the Schwarzschild black hole, setting $Q=0$ in Eq. (\ref{Cp}) and substituting Eqs. (\ref{jTV}), (\ref{jpTV}), and (\ref{Cp}) into Eq. (\ref{JT}), we obtain the JT coefficient $\mu$ as
\begin{align}
\mu&=\f{4r_B\lt[5x^2-8x+4+2(3x-2)\sqrt{1-x}\rt]}{3x^2\lt[5x^2-12x+8+8(x-1)\sqrt{1-x}\rt]}. \n
\end{align}
It is not difficult to find that $\mu$ is positive definite. Moreover, the asymptotic behaviors of $\mu$ are
\begin{align}
\mu&\to\f{4r_+}{3} \qquad (r_B\to r_+), \n\\
\mu&\to\f{5r_B^3}{3r_+^2} \qquad (r_B\to \infty). \n
\end{align}
Hence, $\mu$ tends to a fixed value when $r_B$ approach $r_+$ and monotonically increases at large $r_B$.

Second, for the charged black hole, in the same way, we arrive at
\begin{align}
\mu=\f{2r_B[a(x,y)+3(x^2-y^2)^2(x^2+y^2-2xy^2)]}{3x^2(x^2-y^2)b(x,y)}, \n
\end{align}
where
\begin{align}
a(x,y)&=7x^6 - 16x^5 - 2x^5 y^2 + 8x^4 - 3x^4 y^2 + 32x^3 y^2 + 4x^3 y^4 - 24x^2 y^2 - 39x^2 y^4 \n\\
&\quad + 32x y^4+ 14x y^6 - 13y^6 +4x (3 x^4 - 2 x^3 - 2 x^3 y^2 - 6 x^2 y^2 + 6 x y^2 + 6 x y^4- 5 y^4) \n\\
&\quad \times\sqrt{(1 - x) (1 - y^2/x)}, \n\\
b(x,y)&=5x^4 - 12x^3 - 4x^3 y^2 + 8x^2 + 14x^2 y^2 - 12x y^2 - 4x y^4 + 5y^4 \n\\
&\quad + 8x (x^2 - x - x y^2 + y^2) \sqrt{(1 - x) (1 - y^2/x)}. \n
\end{align}

Similarly, the asymptotic behaviors of $\mu$ are
\begin{align}
\mu&\to\f{4r_+}{3} \qquad (r_B\to r_+), \n\\
\mu&\to\f{r_B^3(5 r_+^2-3 Q^2)}{3 r_+^2 ( r_+^2-Q^2 )} \qquad (r_B\to \infty). \n
\end{align}
Therefore, the charge $Q$ does not affect $\mu$ when $r_B$ approaches $r_+$, but increases $\mu$ at large $r_B$. This is understandable, because the increase of $Q$ is equivalent to the decrease of $r_+$ and the increase of $r_B$ accordingly.

\begin{figure}[h]
\centering\includegraphics[width=0.6\textwidth]{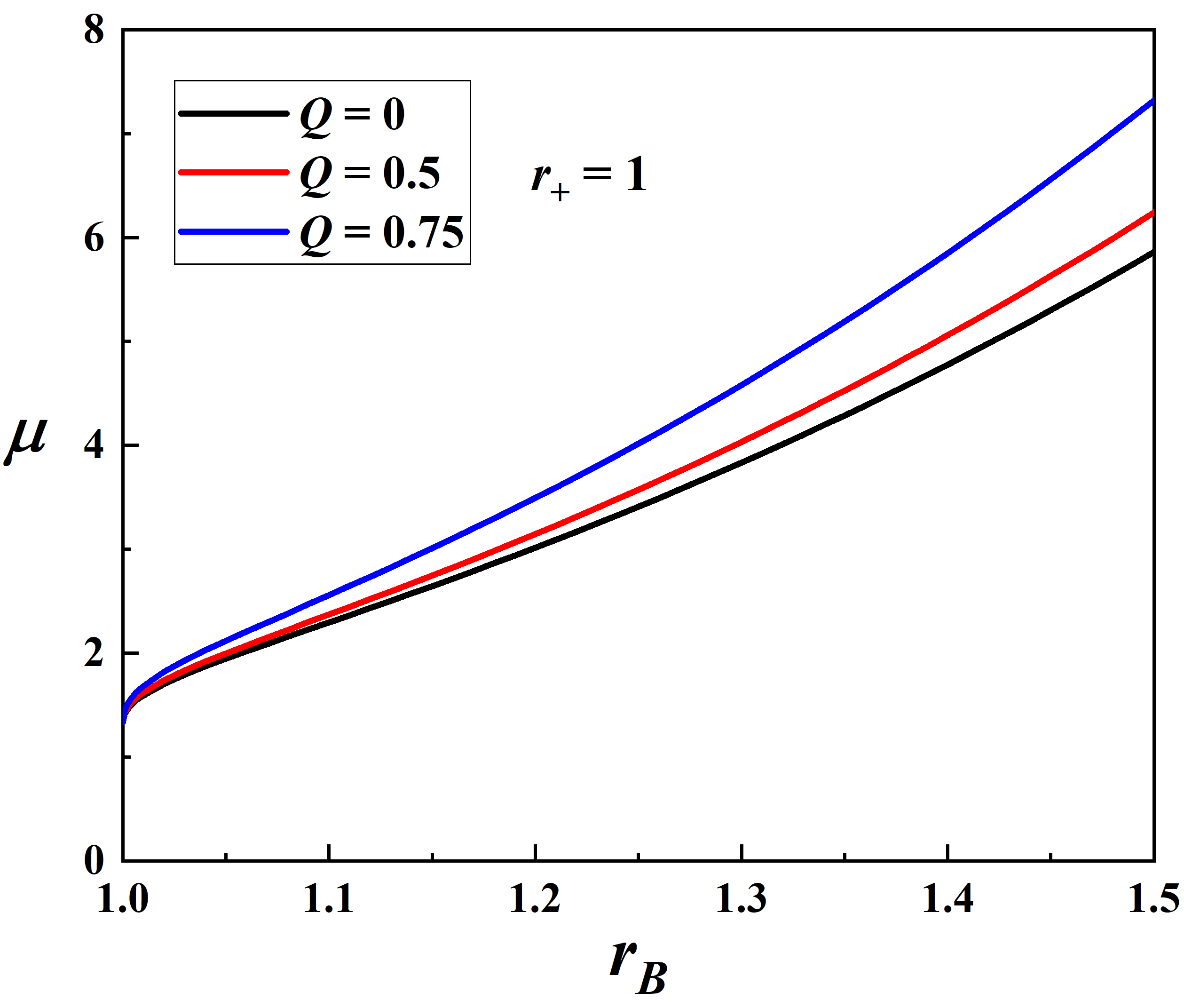}
\caption{The JT coefficient $\mu$ of the black holes in a cavity as a function of the cavity radius $r_B$, with a fixed event horizon radius $r_+= 1$ and different charges $Q=0$, $0.5$, and 0.75, respectively. We observe that $\mu$ is positive definite and monotonically increases with $r_B$ and $Q$. Significantly different from the AdS case, there is only cooling region in the JT effect in a cavity, without inversion temperature or inversion curve.}\label{f:muJT}
\end{figure}

In Fig. \ref{f:muJT}, we plot the JT coefficient $\mu$ of the charged black hole as a function of $r_B$, which is a thermodynamic variable in the extended phase space in a cavity (with a fixed $r_+$ and different $Q$). We clearly observe that $\mu$ is positive definite, meaning that there is only cooling region in the JT effect. This property is remarkably different from that in the AdS space, in which there exists an inversion curve, although there is only a minimum inversion temperature, without a maximum one.

Till now, we are allowed to conclude that there is no inversion temperature or inversion curve in the JT effect of the black holes in the extended phase space in a cavity. However, this is based on the consideration of the JT effect as an isenthalpic process, which is not of much physical meaning in the current situation. Our result is consistent with that in Ref. \cite{Cao:2022hmd}, in which the authors arrived at the similar conclusion as ours mainly by illustration.

\section{JT-like effect of the black holes in a cavity} \label{sec:JTlike}

In this section, we further explore the JT-like effect as a constant-thermal-energy process for the black holes in a cavity and again show that there is no inversion temperature or inversion curve, but the JT-like coefficient $\mu'$ is negative definite at present.

\subsection{Inversion temperature}

Actually, the mathematical derivation in the constant-thermal-energy process is a little simpler, without the complexities from implicit functions. Therefore, from Eqs. (\ref{T}), (\ref{V}), and (\ref{p}), we obtain
\begin{align}
\lt(\f{\p T}{\p p}\rt)_{V,Q}=\f{(\p T/\p r_+)_{V,Q}}{(\p p/\p r_+)_{V,Q}}=\f{(\p T/\p r_+)_{r_B,Q}}{(\p p/\p r_+)_{r_B,Q}}. \label{jd}
\end{align}

First, for the Schwarzschild black hole, setting $Q=0$ in Eqs. (\ref{T}), (\ref{V}), (\ref{p}), and (\ref{jd}), we have
\begin{align}
\f Tp&=\f{2r_B}{x(2-x-2\sqrt{1-x})}, \label{pTS}\\
\lt(\f{\p T}{\p p}\rt)_{V,Q}&=\f{2 r_B(3x-2)}{x^3}. \label{ppTS}
\end{align}
Substituting Eqs. (\ref{pTS}) and (\ref{ppTS}) into Eq. (\ref{fz}), we directly obtain two impossible solutions as $x=0$ and 1 (remember $0<x<1$). Hence, there is no inversion temperature or inversion curve for the Schwarzschild black hole in a cavity even in the JT-like effect.

Second, for the charged black hole, from Eqs. (\ref{T}), (\ref{V}), (\ref{p}), and (\ref{jd}), we have
\begin{align}
\f Tp&=\f{2r_B(r_B^2x^2-Q^2)}{x^2\{r_B^2x[2-x-2\sqrt{f(r_B)}]-Q^2\}}, \label{pTR}\\
\lt(\f{\p T}{\p p}\rt)_{V,Q}&=\f{2 r_B[r_B^4x^3(3x-2)-2r_B^2x(x^2+3x-3)Q^2+(6x-5)Q^4]}{x^2[r_B^4x^4-2r_B^2x^3Q^2+(2x-1)Q^4]}. \label{ppTR}
\end{align}
Substituting Eqs. (\ref{pTR}) and (\ref{ppTR}) into Eq. (\ref{fz}), if there were inversion temperature, we would arrive at
\begin{align}
G(x,y)&=4 x^6 - 8 x^5 - 4 x^5 y^2 +4 x^4+ 16 x^3 y^2+ 8 x^3 y^4 \n\\
&\quad - 12 x^2 y^2 - 24 x^2 y^4 + 16 x y^4 + 4 x y^6 - 4 y^6 \n\\
&\quad +2x (3 x^4 - 2 x^3 - 2 x^3 y^2 - 6 x^2 y^2 + 6 x y^2 + 6 x y^4 - 5 y^4) \sqrt{(1 - x) (1 - y^2/x)} = 0. \n
\end{align}
Again, this equality can never be fulfilled. In fact, $G(x,y)=0$ if and only if $x=1$ or in the limit $y\to x$, consistent with the result for the Schwarzschild black hole; otherwise, $G(x,y)$ is positive definite. Hence, there is no inversion temperature or inversion curve for the charged black hole in a cavity even in the JT-like effect, either.

\subsection{JT-like coefficient}

Now, we calculate the JT-like coefficient $\mu'$, so as to strengthen the above conclusion. It will be interesting to show that $\mu'$ is negative definite in the current situation. Again, the heat capacity at constant volume $C_V$ is
\begin{align}
\f{1}{C_V}=\f 1T\lt(\f{\p T}{\p S}\rt)_{V,Q}=\f{(\p T/\p r_+)_{r_B,Q}}{T\,\dd S/\dd r_+}. \label{CV}
\end{align}

First, for the Schwarzschild black hole, setting $Q=0$ in Eq. (\ref{CV}) and substituting Eqs. (\ref{pTS}), (\ref{ppTS}), and (\ref{CV}) into Eq. (\ref{JTc}), we obtain the JT-like coefficient $\mu'$ as
\begin{align}
\mu'=-\f{3x-2+2(1-x)^{3/2}}{16\pi^2r_B^4x^2(1-x)}. \n
\end{align}
It is evident that $\mu'$ is negative definite. Similarly, the asymptotic behaviors of $\mu'$ are
\begin{align}
\mu'&\to-\f{1}{16\pi^2 r_+^3(r_B-r_+)} \qquad (r_B\to r_+), \n\\
\mu'&\to-\f{3}{64\pi^2r_B^4} \qquad (r_B\to \infty). \n
\end{align}
Hence, $\mu'$ diverges when $r_B$ approaches $r_+$. This is quite different from the situation in the JT effect in Sec. \ref{sec:JT}, where $\mu$ converges to a fixed value $4r_+/3$. Moreover, $\mu'$ monotonically increases at large $r_B$ and approaches 0 when $r_B\to\infty$.

Second, for the charged black hole, in the same way, we obtain
\begin{align}
\mu'&=-\f{3x^4 - 2x^3 - 2x^3 y^2 - 6x^2 y^2 + 6x y^2 + 6x y^4 - 5 y^4} {16\pi^2 r_B^4 x^2(1-x)(x-y^2)(x^2-y^2)}\n\\
&\quad +\f{ (x^4 - x^3 - 2 x^2 y^2 + 3 x y^2 - y^4) \sqrt{(1 - x) (1 - y^2/x)}} {8\pi^2 r_B^4 x^2(1-x)(x-y^2)(x^2-y^2)}, \n
\end{align}
and the asymptotic behaviors of $\mu'$ are
\begin{align}
\mu'&\to-\f{1}{16\pi^2 r_+^3(r_B-r_+)} \qquad (r_B\to r_+), \n\\
\mu'&\to-\f{(r_+^2-Q^2)(3r_+^2-Q^2)}{64\pi^2r_B^4r_+^4} \qquad (r_B\to \infty). \n
\end{align}
Therefore, the charge $Q$ does not affect the divergence of $\mu'$ when $r_B$ approaches $r_+$, and increases $\mu'$ at large $r_B$. In all, the influences of charge on the JT and JT-like coefficients are alike, as explained in Sec. \ref{sec:JT}.

\begin{figure}[h]
\centering\includegraphics[width=0.6\textwidth]{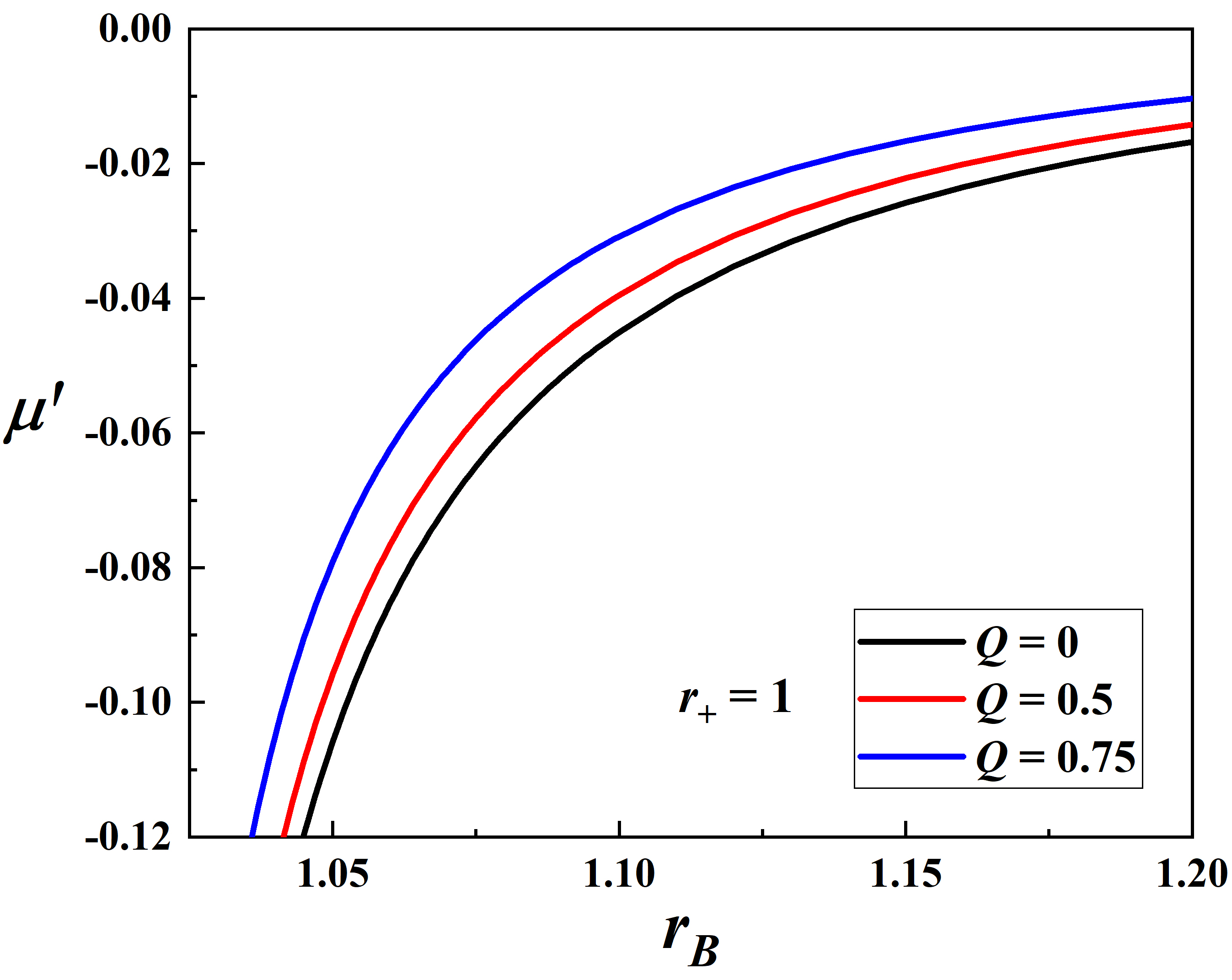}
\caption{The JT-like coefficient $\mu'$ of the black holes in a cavity as a function of the cavity radius $r_B$, with a fixed value of $r_+= 1$ and different values of $Q=0$, $0.5$, and 0.75, respectively. We observe that $\mu'$ is negative definite and monotonically increases with $r_B$ and $Q$. There is only heating region in the JT-like effect in a cavity, without inversion temperature or inversion curve. This behavior is totally different from the JT effect discussed in Sec. \ref{sec:JT} and the JT effect in the AdS space.} \label{f:muJTlike}
\end{figure}

In Fig. \ref{f:muJTlike}, we plot the JT-like coefficient $\mu'$ of the charged black hole as a function of $r_B$ (with a fixed $r_+$ and different $Q$). We easily find that $\mu'$ is negative definite, meaning that there is only heating region in the JT-like effect. This behavior is not only different from the JT effect in the AdS case, but also different from the JT effect as an isenthalpic process in the cavity case, although the latter does not have enough physical essence.

Altogether, we are convinced that, even in the more physically reasonable JT-like effect (i.e., a constant-thermal-energy process), the black holes in the extended phase space in a cavity still have no inversion temperature or inversion curve. In other words, there is only heating region with $\mu'<0$ in the $T$--$V$ plane, just like the cooling region with $\mu>0$ in the $T$--$p$ plane in JT effect in Sec. \ref{sec:JT}.

Last, we spend a few words on the different asymptotic behaviors of $\mu$ and $\mu'$ when $r_B\to\infty$. As shown in Eqs. (\ref{JT}) and (\ref{JTc}), the signs of $\mu$ and $\mu'$ are determined by the EoS of the black hole in a cavity. On the one hand, in the JT effect, for the Schwarzschild black hole, from Eq. (\ref{ddddd}), we have $V\propto T^3$, so
\begin{align}
\mu=\f{1}{C_p}\lt[T\lt(\f{\p V}{\p T}\rt)_{p}-V\rt]=\f{2V}{C_p}>0. \n
\end{align}
For the extremal black hole, from Eq. (\ref{eeeee}), we have $V\propto T^{3/2}$, so
\begin{align}
\mu=\f{1}{C_p}\lt[T\lt(\f{\p V}{\p T}\rt)_{p,Q}-V\rt]=\f{V}{2C_p}>0. \n
\end{align}
Therefore, $\mu$ is positive definite. On the other hand, in the JT-like effect, for the Schwarzschild black hole, from Eq. (\ref{ddddd}), we have $p\propto T$, so
\begin{align}
\mu'=\f{1}{C_V}\lt[p-T\lt(\f{\p p}{\p T}\rt)_{V}\rt]\to 0^-. \n
\end{align}
For the extremal black hole, from Eq. (\ref{eeeee}), we have $p\propto T^2$, so
\begin{align}
\mu'=\f{1}{C_V}\lt[p-T\lt(\f{\p p}{\p T}\rt)_{V,Q}\rt]=-\f{p}{C_V}<0. \n
\end{align}
Hence, $\mu'$ is negative definite. For general black hole solution, except for unnecessary mathematical complexities, the signs of $\mu$ and $\mu'$ will not change. These observations explain the interesting asymptotic behaviors of $\mu$ and $\mu'$ when $r_B\to\infty$ in different extended phase spaces.

\section{Conclusion}\label{sec:con}

Black hole thermodynamics in the extended phase space has attracted increasing attention recently. One of the basic motivations is to restore the absent $p$--$V$ term in the first law of thermodynamics, and this can be achieved via two complementary methods: to confine the black hole in the AdS space or to enclose it in a cavity with the radius $r_B$. In these two extended phase spaces, a thermodynamically stable black hole has an EoS analogous to that of a non-ideal fluid and can thus exhibit abundant thermodynamic behaviors. In our present work, a typical thermodynamic process (the JT effect) and a relevant process (the JT-like effect) are investigated in detail for the black holes in a cavity.

This work is a succession and comparison with our previous study on the JT effect as an isenthalpic process in the AdS space \cite{Zhao:2018kpz}, in which we systematically explored the JT coefficient, inversion temperature, inversion curve, and isenthalpic curve, respectively. However, an isenthalpic process is not of much physical essence for the black holes in a cavity, because in this case, the thermodynamic variable in the first law is thermal energy instead of enthalpy. Consequently, in this work, we investigate not only the JT effect as an isenthalpic (constant-mass) process, but also the JT-like effect as a constant-thermal-energy process. It is found that in both cases, there is no inversion temperature or inversion curve. In the former, there is only cooling region with a positive definite JT coefficient $\mu$; on the contrary, in the latter, there is only heating region with a negative definite JT-like coefficient $\mu'$. The different signs and asymptotic behaviors of $\mu$ and $\mu'$ when $r_B\to \infty$ are basically determined by the EoS of the black holes. All these characters are in contradiction with those in the AdS space, in which there exist inversion temperature and inversion curve. In fact, our present work reflects only one of the various differences between the two extended phase spaces, like those discovered in our earlier studies on the Hawking--Page phase transition \cite{zwb, Su:2021jto}. All these observations indicate that, despite the remarkable similarities in the two extended phase spaces, there still remain some notable discrepancies between them that are sensitive to their specific boundary conditions and thus deserve further reevaluation.

\acknowledgments

We are very grateful to Peng Wang, Wen-Bo Zhao, and Ze-Wei Zhao for fruitful discussions.


\begin{thebibliography}{99}
\bibitem{law}
J.M. Bardeen, B. Carter, S.W. Hawking, \emph{The four laws of black hole mechanics}, \emph{Commun. Math. Phys.} {\bf 31} (1973) 161.

\bibitem{rad}
S. W. Hawking, \emph{Particle creation by black holes}, \emph{Commun. Math. Phys.} {\bf 43} (1975) 199.

\bibitem{Kastor}
D. Kastor, S. Ray, J. Traschen, \emph{Enthalpy and the mechanics of AdS black holes},
\emph{Class. Quant. Grav.}{\bf 26} (2009) 195011, \arxth{0904.2765}. 

\bibitem{Dolan}
B. P. Dolan, \emph{Pressure and volume in the first law of black hole thermodynamics},
\emph{Class. Quant. Grav.} {\bf 28} (2011) 235017, \arxgr{1106.6260}. 

\bibitem{KM}
D. Kubiz\v{n}\'{a}k, R. B. Mann, \emph{$P$--$V$ criticality of charged AdS black holes},
\emph{J. High Energy Phys.} {\bf 07} (2012) 033, \arxth{1205.0559}. 

\bibitem{rev}
D. Kubiz\v{n}\'{a}k, R. B. Mann, M. Teo, \emph{Black hole chemistry: thermodynamics with Lambda},
\emph{Class. Quant. Grav.} {\bf 34} (2017) 063001, \arxth{1608.06147}. 

\bibitem{York}
J. W. York, \emph{Black hole thermodynamics and the Euclidean Einstein action}, \emph{Phys. Rev. D} {\bf 33} (1986) 2092.

\bibitem{Braden}
H. W. Braden, J. D. Brown, B. F. Whiting, J. W. York, \emph{Charged black hole in a grand canonical ensemble},
\emph{Phys. Rev. D} {\bf 42} (1990) 3376.

\bibitem{Parentani:1994wr}
R. Parentani, J. Katz, I. Okamoto, \emph{Thermodynamics of a black hole in a cavity},
\emph{Class. Quant. Grav.} {\bf 12} (1995) 1663, \Arxgr{9410015}. 

\bibitem{Gregory:2001bd}
J. P. Gregory, S. F. Ross, \emph{Stability and the negative mode for Schwarzschild in a finite cavity},
\emph{Phys. Rev. D} {\bf 64} (2001) 124006, \Arxth{0106220}. 

\bibitem{Carlip}
S. Carlip, S. Vaidya, \emph{Phase transitions and critical behavior for charged black holes},
\emph{Class. Quant. Grav.} {\bf 20} (2003) 3827, \Arxgr{0306054}. 

\bibitem{Lundgren}
A. P. Lundgren, \emph{Charged black hole in a canonical ensemble},
\emph{Phys. Rev. D} {\bf 77} (2008) 044014, \Arxgr{0612119}.

\bibitem{Akbar:2004ke}
M. M. Akbar, \emph{Schwarzschild-anti de Sitter within an isothermal cavity: Thermodynamics, phase transitions and the Dirichlet problem},
\emph{Phys. Rev. D} {\bf 82} (2010) 064001, \Arxth{0401228}. 

\bibitem{Emparan:2012be}
R. Emparan, M. Martinez, \emph{Black branes in a box: hydrodynamics, stability, and criticality},
\emph{J. High Energy Phys.} {\bf 07} (2012) 120, \arxth{1205.5646}. 

\bibitem{Dolansss}
S. R. Dolan, S. Ponglertsakul, E. Winstanley, \emph{Stability of black holes in Einstein--charged scalar field theory in a cavity},
\emph{Phys. Rev. D} {\bf 92} (2015) 124047, \arxgr{1507.02156}. 

\bibitem{27}
N. Sanchis-Gual, J. C. Degollado, P. J. Montero, J. A. Font, C. Herdeiro,
\emph{Explosion and final state of an unstable Reissner--Nordstr\"{o}m black hole},
\emph{Phys. Rev. Lett.} {\bf 116} (2016) 141101, \arxgr{1512.05358}. 

\bibitem{Ponglertsakul1}
S. Ponglertsakul, S. R. Dolan, E. Winstanley, \emph{Stability of gravitating charged-scalar solitons in a cavity},
\emph{Phys. Rev. D} {\bf 94} (2016) 024031, \arxgr{1604.01132}. 

\bibitem{28}
N. Sanchis-Gual, J. C. Degollado, C. Herdeiro, J. A. Font, P. J. Montero,
\emph{Dynamical formation of a Reissner--Nordstr\"{o}m black hole with scalar hair in a cavity},
\emph{Phys. Rev. D} {\bf 94} (2016) 044061, \arxgr{1607.06304}. 

\bibitem{29}
P. Basu, C. Krishnan, P. N. Bala Subramanian, \emph{Hairy black holes in a box},
\emph{J. High Energy Phys.} {\bf 11} (2016) 041, \arxth{1609.01208}. 

\bibitem{Ponglertsakul2}
S. Ponglertsakul, E. Winstanley, \emph{Effect of scalar field mass on gravitating charged scalar solitons and black holes in a cavity},
\emph{Phys. Lett. B} {\bf 87} (2017) 764, \arxgr{1610.00135}. 

\bibitem{Sanchis-Gualhhh}
N. Sanchis-Gual, J. C. Degollado, J. A. Font, C. Herdeiro, E. Radu,
\emph{Dynamical formation of a hairy black hole in a cavity from the decay of unstable solitons},
\emph{Class. Quant. Grav.} {\bf 34} (2017) 165001, \arxgr{1611.02441}. 

\bibitem{31}
Y. Peng, \emph{Studies of a general flat space/boson star transition model in a box through a language similar to holographic superconductors},
\emph{J. High Energy Phys.} {\bf 07} (2017) 042, \arxth{1705.08694}. 

\bibitem{30}
Y. Peng, B. Wang, Y. Liu,
\emph{On the thermodynamics of the black hole and hairy black hole transitions in the asymptotically flat spacetime with a box},
\emph{Eur. Phys. J. C} {\bf 78} (2018) 176, \arxth{1708.01411}. 

\bibitem{Dias11}
O. J. C. Dias, R. Masachs, \emph{Charged black hole bombs in a Minkowski cavity},
\emph{Class. Quant. Grav.} {\bf 35} (2018) 184001, \arxgr{1801.10176}. 

\bibitem{Dias12}
O. J. C. Dias, R. Masachs, \emph{Evading no-hair theorems: hairy black holes in a Minkowski box},
\emph{Phys. Rev. D} {\bf 97} (2018) 124030, \arxgr{1802.01603}. 

\bibitem{35}
F. Simovic, R. B. Mann, \emph{Critical phenomena of charged de Sitter black holes in cavities},
\emph{Class. Quant. Grav.} {\bf 36} (2019) 014002, \arxgr{1807.11875}. 

\bibitem{Wang:2019kxp}
P. Wang, H. Wu, H. Yang, \emph{Thermodynamics and phase transition of a nonlinear electrodynamics black hole in a cavity},
\emph{J. High Energy Phys.} {\bf 07} (2019) 002, \arxgr{1901.06216}. 

\bibitem{36}
F. Simovic, R. B. Mann, \emph{Critical phenomena of Born--Infeld-de Sitter black holes in cavities},
\emph{J. High Energy Phys.} {\bf 05} (2019) 136, \arxgr{1904.04871}. 

\bibitem{Wang:2019urm}
P. Wang, H. Yang, S. Ying, \emph{Thermodynamics and phase transition of a Gauss--Bonnet black hole in a cavity},
\emph{Phys. Rev. D} {\bf 101} (2020) 064045, \arxgr{1909.01275}. 

\bibitem{Wang:2019cax}
P. Wang, H. Wu, H. Yang, \emph{Thermodynamic geometry of AdS black holes and black holes in a cavity},
\emph{Eur. Phys. J. C} {\bf 80} (2020) 216, \arxgr{1910.07874}. 

\bibitem{37}
S. Haroon, R. A. Hennigar, R. B. Mann, F. Simovic, \emph{Thermodynamics of Gauss--Bonnet-de Sitter black holes},
\emph{Phys. Rev. D} {\bf 101} (2020) 084051, \arxgr{2002.01567}. 

\bibitem{Wang:2020osg}
P. Wang, H. Wu, S. Ying, \emph{Validity of thermodynamic laws and weak cosmic censorship for AdS black holes and black holes in a cavity},
\emph{Chin. Phys. C} {\bf 45} (2021) 055105, \arxgr{2002.12233}. 

\bibitem{Tzikas}
A. G. Tzikas, \emph{Regular black holes in isothermal cavity},
\emph{Phys. Lett. B} {\bf 819} (2021) 136426, \arxth{2012.03349}.

\bibitem{zwb}
W.-B. Zhao, G.-R.~Liu, N. Li, \emph{Hawking Page phase transitions of the black holes in a cavity},
\emph{Eur. Phys. J. Plus} {\bf 136} (2021) 981, \arxgr{2012.13921}.

\bibitem{Yao:2021zid}
F. Yao, \emph{Scalarized Einstein--Maxwell-scalar black holes in a cavity},
\emph{Eur. Phys. J. C} {\bf 81} (2021) 1009, \arxgr{2107.12039}.

\bibitem{ElMoumni:2021woq}
H. El Moumni, J. Khalloufi, \emph{Nonlinear--Maxwell--Yukawa de-Sitter black hole thermodynamics in a cavity: I Canonical ensemble},
\emph{Nucl. Phys. B} {\bf 973} (2021) 115593. 

\bibitem{Su:2021jto}
B.-Y. Su, N. Li, \emph{On the dual relation in the Hawking Page phase transition of the black holes in a cavity},
\emph{Nucl. Phys. B} {\bf 979} (2022) 115782, \arxgr{2105.06670}.

\bibitem{Wang:2021llu}
P. Wang, F. Yao, \emph{Thermodynamic geometry of black holes enclosed by a cavity in extended phase space},
\emph{Nucl. Phys. B} {\bf 976} (2022) 115715, \arxgr{2107.14640}.

\bibitem{Feng:2021zsq}
Z. W. Feng, X. Zhou, S. Q. Zhou, S. Z. Yang,
\emph{Quantum corrections to the thermodynamics and phase transition of a black hole surrounded by a cavity in the extended phase space},
\emph{Commun. Theor. Phys.} {\bf 74} (2022) 085403, \arxgr{2109.13667}.

\bibitem{Huang:2021eby}
Y. Huang, J. Tao, \emph{Thermodynamics and phase transition of BTZ black hole in a cavity},
\emph{Nucl. Phys. B} {\bf 982} (2022) 115881, \arxgr{2112.13249}.

\bibitem{Cao:2022hmd}
Y. Cao, H. Feng, J. Tao, Y. Xue, \emph{Black Holes in a Cavity: Heat engine and Joule--Thomson Expansion},
\emph{Gen. Rel. Grav.} {\bf 54} (2022) 105, \arxgr{2201.07584}.

\bibitem{Huang:2022agr}
Y. Huang, J. Tao, P. Wang, S. Ying,
\emph{Thermodynamics and phase transitions of a Kerr--Newman black hole in a cavity}, \emph{Eur. Phys. J. Plus} {\bf 138} (2023) 3265, \arxgr{2206.10055}.

\bibitem{Wang}
P. Wang, H. Wu, H. Yang, F. Yao, \emph{Extended phase space thermodynamics for black holes in a cavity},
\emph{J. High Energy Phys.} {\bf 09} (2020) 154, \arxgr{2006.14349}. 

\bibitem{okcu1}
\"{O}. \"{O}kc\"{u}, E. Ayd{\i}ner, \emph{Joule--Thomson expansion of the charged AdS black holes},
\emph{Eur. Phys. J. C} {\bf 77} (2017) 24, \arxgr{1611.06327}. 

\bibitem{okcu2}
\"{O}. \"{O}kc\"{u}, E. Ayd{\i}ner, \emph{Joule--Thomson expansion of Kerr--AdS black holes},
\emph{Eur. Phys. J. C} {\bf 78} (2018) 123, \arxgr{1709.06426}. 

\bibitem{Zhao:2018kpz}
Z.-W. Zhao, Y.-H. Xiu, N.~Li,
\emph{Throttling process of the Kerr--Newman--anti-de Sitter black holes in the extended phase space},
\emph{Phys. Rev. D} {\bf 98} (2018) 124003, \arxgr{1805.04861}. 

\bibitem{Ghaffarnejad:2018exz}
H. Ghaffarnejad, E. Yaraie, M. Farsam, \emph{Quintessence Reissner Nordstr\"om Anti de Sitter Black Holes and Joule Thomson effect},
\emph{Int. J. Theor. Phys.} {\bf 57} (2018) 1671, 
\arxgr{1802.08749}.

\bibitem{Mo:2018rgq}
J.-X. Mo, G.-Q. Li, S.-Q. Lan, X.-B. Xu, \emph{Joule--Thomson expansion of $d$-dimensional charged AdS black holes},
\emph{Phys. Rev. D} {\bf 98} (2018) 124032, 
\arxgr{1804.02650}.

\bibitem{Chabab:2018zix}
M. Chabab, H. El Moumni, S. Iraoui, K. Masmar, S. Zhizeh, \emph{Joule--Thomson Expansion of RN--AdS Black Holes in $f(R)$ gravity},
\emph{Lett. Energy High Phys.} {\bf 02} (2018) 05, 
\arxgr{1804.10042}.

\bibitem{Mo:2018qkt}
J.-X. Mo, G.-Q. Li, \emph{Effects of Lovelock gravity on the Joule--Thomson expansion},
\emph{Classical Quantum Gravity} {\bf 37} (2020) 045009, 
\arxgr{1805.04327}.

\bibitem{Lan:2018nnp}
S.-Q. Lan, \emph{Joule--Thomson expansion of charged Gauss--Bonnet black holes in AdS space},
\emph{Phys. Rev. D} {\bf 98} (2018) 084014, 
\arxgr{1805.05817}.


\bibitem{Kuang:2018goo}
X.-M. Kuang, B. Liu, A. \"Ovg\"un, \emph{Nonlinear electrodynamics AdS black hole and related phenomena in the extended thermodynamics},
\emph{Eur. Phys. J. C} {\bf 78} (2018) 840, 
\arxgr{1807.10447}.

\bibitem{Cisterna:2018jqg}
A. Cisterna, S.-Q. Hu, X.-M. Kuang, \emph{Joule--Thomson expansion in AdS black holes with momentum relaxation},
\emph{Phys. Lett. B} {\bf 797} (2019) 134883, 
\arxgr{1808.07392}.

\bibitem{Haldar:2018cks}
A. Haldar R. Biswas, \emph{Joule--Thomson expansion of five-dimensional Einstein--Maxwell--Gauss--Bonnet--AdS black holes},
\emph{Europhys. Lett.} {\bf 123} (2018) 40005. 

\bibitem{Li:2019jcd}
C. Li, P. He, P. Li, J.-B. Deng, \emph{Joule--Thomson expansion of the Bardeen--AdS black holes}, \emph{Gen. Relativ. Gravit.} {\bf 52} (2020) 50, 
\arxgr{1904.09548}.


\bibitem{MahdavianYekta:2019dwf}
D. Mahdavian Yekta, A. Hadikhani, \"O. \"Okc\"u, \emph{Joule--Thomson expansion of charged AdS black holes in Rainbow gravity},
\emph{Phys. Lett. B} {\bf 795} (2019) 521, 
\arxth{1905.03057}.

\bibitem{Nam:2019zyk}
C. H. Nam, \emph{Heat engine efficiency and Joule--Thomson expansion of nonlinear charged AdS black hole in massive gravity},
\emph{Gen. Relativ. Gravit.} {\bf 53} (2021) 30, 
\arxgr{1906.05557}.


\bibitem{Lan:2019kak}
S.-Q. Lan, \emph{Joule--Thomson expansion of neutral AdS black holes in massive gravity},
\emph{Nucl. Phys. B} {\bf 948} (2019) 114787. 


\bibitem{Sadeghi:2020bon}
J. Sadeghi, R. Toorandaz, \emph{Joule--Thomson expansion of hyperscaling violating black holes with spherical and hyperbolic horizons},
\emph{Nucl. Phys. B} {\bf 951} (2020) 114902. 

\bibitem{Rajani:2020mdw}
K. V. Rajani, C. L. A. Rizwan, A. Naveena Kumara, M. S. Ali, D. Vaid,
\emph{Joule--Thomson expansion of regular Bardeen AdS black hole surrounded by static anisotropic matter field},
\emph{Phys. Dark Univ.} {\bf 32} (2021) 100825, 
\arxgr{2002.03634}.

\bibitem{Nam:2020gud}
C. H. Nam, \emph{Effect of massive gravity on Joule--Thomson expansion of the charged AdS black hole}, \emph{Eur. Phys. J. Plus} {\bf 135} (2020) 259. 


\bibitem{Guo:2020zcr}
S. Guo, Y. Han, G.-P. Li,
\emph{Thermodynamic of the charged AdS black holes in Rastall gravity: $P$--$V$ critical and Joule--Thomson expansion},
\emph{Mod. Phys. Lett. A} {\bf 35} (2020) 2050113. 

\bibitem{Hegde:2020xlv}
K. Hegde, A. Naveena Kumara, C. L. A. Rizwan, K. M. Ajith, M. S. Ali,
\emph{Thermodynamics, Phase Transition and Joule--Thomson Expansion of novel $4$-$D$ Gauss Bonnet AdS Black Hole},
\arxgr{2003.08778}.

\bibitem{Guo:2020qxy}
S. Guo, Y. Han, G.-P. Li,
\emph{Joule--Thomson expansion of a specific black hole in $f(R)$ gravity coupled with Yang--Mills field},
\emph{Class. Quant. Grav.} {\bf 37} (2020) 085016. 


\bibitem{Bi:2020vcg}
S. Bi, M. Du, J. Tao, F. Yao, \emph{Joule--Thomson expansion of Born--Infeld AdS black holes},
\emph{Chin. Phys. C} {\bf 45} (2021) 025109, 
\arxgr{2006.08920}.


\bibitem{Silva:2021qtw}
G. V. Silva, V. B. Bezerra, J. P. M. Gra\c{c}a, I. P. Lobo,
\emph{Joule--Thomson expansion in charged AdS black hole surrounded by a cosmological fluid in Rainbow Gravity},
\emph{Mod. Phys. Lett. A} {\bf 36} (2021) 2150278. 

\bibitem{Mirza:2021kvi}
B. Mirza, F. Naeimipour, M. Tavakoli,
\emph{Joule--Thomson Expansion of the Quasitopological Black Holes},
\emph{Front. in Phys.} {\bf 9} (2021) 33, 
\arxgr{2105.05047}.


\bibitem{Zhang:2021raw}
M. Zhang, C.-M. Zhang, D.-C. Zou, R.-H. Yue,
\emph{$P$--$V$ criticality and Joule--Thomson expansion of Hayward--AdS black holes in $4D$ Einstein--Gauss--Bonnet gravity},
\emph{Nucl. Phys. B} {\bf 973} (2021) 115608, 
\arxth{2102.04308}.

\bibitem{Liang:2021elg}
J. Liang, W. Lin, B. Mu, \emph{Joule--Thomson expansion of the torus-like black hole},
\emph{Eur. Phys. J. Plus} {\bf 136} (2021) 1169, 
\arxgr{2103.03119}.

\bibitem{Liang:2021xny}
J. Liang, B. Mu, P. Wang, \emph{Joule--Thomson expansion of lower-dimensional black holes},
\emph{Phys. Rev. D} {\bf 104} (2021) 124003, 
\arxgr{2104.08841}.

\bibitem{Graca:2021izb}
J. P. M. Gra\c{c}a, E. F. Capossoli, H. Boschi-Filho,
\emph{Joule--Thomson expansion for quantum corrected AdS--Reissner--Nordstrom black holes in Kiselev spacetime},
\emph{Phys. Rev. D} {\bf 107} (2023) 024045, \arxgr{2105.04689}.

\bibitem{Yin:2021akt}
R. Yin, J. Liang, B. Mu,
\emph{Joule--Thomson expansion of Reissner--Nordstr\"om--Anti-de Sitter black holes with cloud of strings and quintessence},
\emph{Phys. Dark Univ.} {\bf 34} (2021) 100884, 
\arxgr{2105.09173}.

\bibitem{Zhang:2021kha}
C.-M. Zhang, M. Zhang, D.-C. Zou,\emph{Joule--Thomson expansion of Born--Infeld AdS black holes in consistent $4D$ Einstein--Gauss--Bonnet gravity},
\emph{Mod. Phys. Lett. A} {\bf 37} (2022) 2250063, 
\arxth{2106.00183}.

\bibitem{Biswas:2021uop}
A. Biswas, \emph{Joule--Thomson expansion of AdS black holes in Einstein Power--Yang--mills gravity},
\emph{Phys. Scripta} {\bf 96} (2021) 125310, 
\arxgr{2106.11066}.

\bibitem{Graca:2021ker}
J. P. M. Gra\c{c}a, E. F. Capossoli, H. Boschi-Filho, \emph{Joule--Thomson expansion for noncommutative uncharged black holes},
\emph{Europhys. Lett.} {\bf 135} (2021) 41002, 
\arxth{2107.05781}.

\bibitem{Meng:2021cgb}
Y. Meng, B.-B. Chen, J. Tang, \emph{Cooling heating phase transition of the Euler Heisenberg--AdS black hole},
\emph{Mod. Phys. Lett. A} {\bf 36} (2021) 2150165. 

\bibitem{Abdusattar:2021wfv}
H. Abdusattar, S.-B. Kong, W.-L. You, H. Zhang, Y.-P. Hu,
\emph{Thermodynamic Equation of State and Joule--Thomson Expansion for a FRW Universe}, \emph{J. High Energy Phys.} {\bf 12} (2022) 168,
\arxgr{2108.09407}.

\bibitem{Xing:2021gpn}
J.-T. Xing, Y. Meng, X.-M. Kuang, \emph{Joule--Thomson expansion for hairy black holes},
\emph{Phys. Lett. B} {\bf 820} (2021) 136604. 

\bibitem{Zarepour:2021xmz}
S. Zarepour, \emph{Holographic Joule--Thomson expansion in lower dimensions},
\emph{Phys. Scripta} {\bf 96} (2021) 125011, 
\arxth{2110.13829}.

\bibitem{Assrary:2022uiu}
M. Assrary, J. Sadeghi, M. E. Zomorrodian, \emph{The effect of nonlinear electrodynamics on Joule--Thomson expansion of a $5$-dimensional charged AdS black hole in Einstein--Gauss--Bonnet gravity},
\emph{Nucl. Phys. B} {\bf 977} (2022) 115727. 

\bibitem{Kruglov:2022lnc}
S. I. Kruglov, \emph{Nonlinearly charged AdS black holes, extended phase space thermodynamics and Joule--Thomson expansion},
\emph{Annals Phys.} {\bf 441} (2022) 168894, 
\arxph{2208.13662}

\bibitem{Yin:2022mcy}
R. Yin, J. Liang, Y. Song, Y. He, B. Mu,
\emph{Joule--Thomson expansion of $d$-dimensional charged AdS black holes with cloud of strings and quintessence},
\arxgr{2204.08314}.

\bibitem{Barrientos:2022uit}
J. Barrientos, J. Mena, \emph{Joule--Thomson expansion of AdS black holes in quasi-topological electromagnetism},
\emph{Phys. Rev. D} {\bf 106} (2022) 044064, \arxgr{2206.06018}.

\bibitem{Hui:2022ubj}
S. Hui, B. Mu, J. Tao, \emph{Joule--Thomson expansion of the lower-dimensional black hole in rainbow gravity},
\arxgr{2207.01467}.


\bibitem{Smarr}
L. Smarr, \emph{Mass formula for Kerr black holes},
\emph{Phys. Rev. Lett.} {\bf 30} (1973) 71. 

\end{thebibliography}
\end{document}